# Microcontroller Based Portable Measurement System for GaN and SiC Devices Characterization

Alberto Vella[1], Giuseppe Galioto[1] and Giuseppe Costantino Giaconia[1]

[1] University of Palermo,
Department of Engineering, Viale delle scienze Ed. 9, 90128 Palermo, Italy
{alberto.vella, giuseppe.galioto, costantino.giaconia}@unipa.it

**Abstract.** The aim of this work is to design and implement an embedded system capable to characterize some relevant figures of merit of Gallium Nitride and Silicon Carbide transistors in a wide range of frequencies. In particular, the designed system is focused on measuring the parameters involved in both the power loss phenomena and the reliability of the device during switching operations. Both the employment of a low-cost microcontroller unit and the equivalent-time sampling technique contributed to make the measurement system flexible, affordable and capable of enhanced sampling performance. As a result, different GaN and SiC devices were compared, in order to characterize the behavior of the measured quantities with respect to the switching frequency.

## 1  Introduction

Reducing power losses in conversion systems has always been a critical issue in most power electronics applications. Nowadays, systems with low losses, and high efficiency, allow for a better power conversion with enhanced performance and lower costs. As explained in [1], main losses in power electronics systems are caused by the different phenomena related to each switching device (HEMT, MOSFET, BJT, IGBT) but all of them can be seen as the sum of several terms:

$$P_L = P_{on} + P_s + P_{off} \ . \tag{1}$$

Where, $P_L$ is the total power loss, $P_s$ represents commutation losses due to switching operations, $P_{on}$ is the conduction loss during the on-state and $P_{off}$ is the leakage term collecting losses related to the off-state.

Since the drain current is almost zero (~µA), when a device is off, the related $P_{off}$ term can be generally neglected, then the total power loss can be well approximated with the sum of switching and conduction losses:

$$P_L \cong P_{on} + P_s \ . \tag{2}$$

Conduction losses occur during the transistor's on-state and can be expressed as:

$$P_{on} = R_{DSon} I_0^2 \frac{t_{on}}{T_S} . \tag{3}$$

In which $R_{DSon}$ is the on-resistance of the switching device, $I_0$ is the on-state current, $t_{on}$ is the on-state time and $T_S$ is the switching period. So, as it can be seen from the previous equation, conduction losses depend on the on-resistance and, consequently, they can be minimized using switching devices with low $R_{DSon}$, like SiC MOSFET or GaN HEMT because the theoretical limit of the on-resistance for these devices is much lower if compared with the limit of Silicon devices.

Switching power losses occur when, during the on- or off- state transition, the drain-source voltage $V_d$ and the drain current $I_0$ are simultaneously different from zero.

$$P_s = \frac{1}{2} V_d I_0 f_s (t_{c(on)} + t_{c(off)}) . \tag{4}$$

It is possible to notice that $P_s$ linearly depends, as well as on the switching frequency $f_s$, on the switching time $t_{c(on)}$ and $t_{c(off)}$, which are obviously related to the parasitic capacitance of the switching device. For this reason, the input capacitance $C_{iss}$ is a common parameter to be considered and its value can strongly affect power conversion performances. So, it is desirable to have a device with small input capacitance, in order to reduce the switching times and, therefore, switching power losses.

Furthermore, transistors' reliability can strongly be affected by threshold voltage instability phenomena. A negative threshold voltage shift can bring to unwanted turn-on of the device, while a positive shift can make switching times longer and affect the on-resistance, so this phenomenon can increase the device's power losses if there is an increase of $R_{DSon}$, this will particularly bring to larger conduction losses.

All these considerations lead to the necessity of a direct measurement of some important figures of merit of the switching devices, such as the on-resistance, the input capacitance and the threshold voltage variations. It is also important to evaluate the behavior of these quantities with respect to the switching frequency, in order to characterize the device in a wide range of frequencies.

## 2 Embedded System Design

### 2.1 System Requirements

To perform all the measurements of interest, the designed system must satisfy some requirements due to the number of quantities to be measured and to the nature of the involved signals. The system, in fact, must be able to sample at the same time at least two different signals between the gate voltage, the drain current and the drain voltage. All the measurement must be performed in a wide range of switching frequency. In particular, the selected frequency range is between 10kHz and 1MHz, in order to investigate the behavior of high frequency devices like the Gallium Nitride HEMTs.

## 2.2 Designed System

In order to satisfy all the requirements, most importantly the frequency requirements, the STM32H743ZI microcontroller has been selected. This choice was due to the relatively high clock frequency (480MHz), the presence of three independent ADCs and the presence of a high resolution timer (HRTIM), which can use the microcontroller clock as a clock source.

The threshold voltage was extracted using the *constant current* method explained in [2]. So, the threshold voltage is calculated as the gate voltage corresponding to a predetermined drain current. To select the constant drain current $I_C$, the absolute drain current measurement uncertainty, $u_{I_D}$, is considered. The selected drain current value is chosen equal to 100 times $u_{I_D}$. So, during the OFF-ON transition the constant current threshold is $I_C$, while during the ON-OFF transition the current threshold is the ON-state drain current value, lowered by a quantity equal to $I_C$.

The drain-source on-resistance is measured starting from the drain current $I_D$ and the drain voltage $V_D$, since $R_{DSon}$ is equal to the ratio between these two quantities. A gate signal with 90% duty cycle is used in order to switch the device and during conduction state, an average value of $V_D$ and $I_D$ samples is calculated, and these averaged values are used to extract $R_{DSon}$ measurements.

The input capacitance is instead calculated using the following equation, which relates this quantity to the rising time of the gate voltage.

$$C_{iss} = \frac{\tau}{R_G} . \qquad (5)$$

Where $\tau$ is the time constant of the gate voltage transient and $R_G$ is the external gate resistance.

In order to sample signals which frequency can reach 1MHz using a microcontroller system, normal techniques would bring to use ADC with very high sampling frequency, while we wanted to keep is relatively low. For this reason, the *equivalent-time sampling* method has been implemented. This sampling technique can be used only if the measured signal is periodic. It consists in sampling the input signal with a sampling period equal to the signal period plus an additional time defined as $T_{ET} = T/n$, where $T$ is the input signal period and $n$ is the desired number of samples. So, the sampling period can be written as:

$$T_S = T + T_{ET} . \qquad (6)$$

Exploiting the equivalent-time sampling technique allows to sample a signal with a sampling frequency even slightly lower than the signal's frequency.

## 3 Experimental results

The experimental tests carried out in order to characterize GaN and SiC devices were performed on the following three products: a 900V 15A GaN-Cascode, a 650V 15A GaN E-mode and a 900V 11.5A SiC MOSFET.

The measurements for GaN devices extend to 1 MHz, while for SiC MOSFET the maximum achieved switching frequency is equal to 200 kHz. Moreover, the bias conditions of the tested devices are different: for GaN Cascode and E-mode the power supply is 60V and 0.4A, while for SiC MOSFET is 50V and 2A. This choice is due to the different characteristics of SiC devices, whose saturation needs a higher drain current level. Furthermore, SiC device is driven with a 0-15V gate voltage, while for GaN devices a driving voltage of ±5V is already enough to push an pop from saturation. In Table 1 the absolute measurement uncertainties of the considered quantities are shown. Since for the threshold voltage extraction the constant current method is employed, the absolute uncertainty cannot be calculated and so the relative one is presented.

**Table 1.** Measurement absolute uncertainty of the devices' figures of merit

| Figure of merit | Uncertainty | Type of uncertainty |
|---|---|---|
| $C_{iss}$ | 35 pF (worst case) | Absolute |
| $R_{DSon}$ | 0.022 mΩ | Absolute |
| $\Delta V_{th}$ | 0.007 % | Relative |

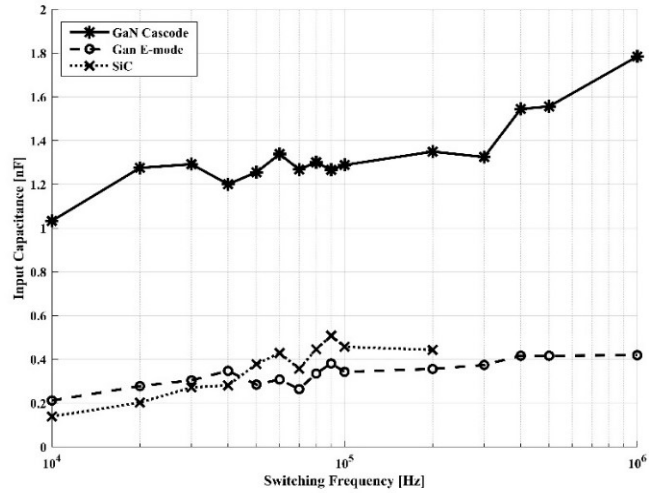

**Fig. 1.** Input capacitance vs switching frequency

As shown in Fig. 1, the input capacitance of all the tested devices approximately doubled their low frequency value in the considered range of frequencies. The physical phenomenon behind this behavior resides in the electron trapping occurring within the switching period.

In Fig. 2 the extracted relation between $R_{DSon}$ and the switching frequency is shown. The measured values are referred to the on-resistance at a switching frequency of 10 kHz. This choice is motivated by the difference in the $R_{DSon}$ values between the devices under test, allowing for a better visualization of the frequency behavior. In SiC device, the variation is minimal, while GaN devices show a more consistent increase in the high-frequency region, approximately 25-30 % at 1 MHz.

The last figure of merit that is characterized is the threshold voltage variation during a switching period, as shown in Fig. 3. The post-processed values of $\Delta V_{th}$ are calculated as difference between the positive threshold voltage variation during the on-state and the negative one during the off-state. GaN Cascode shows an almost null variation of its threshold voltage as a function of switching frequency. GaN E-mode's threshold voltage variation increases at high switching frequencies of hundreds of mV. Finally, the highest variation occurs for the SiC device, whose threshold voltage increases over 4V at increasing switching frequencies.

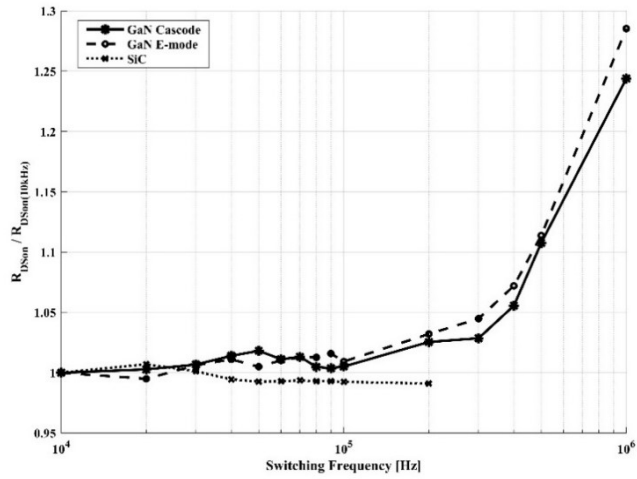

**Fig. 2.** $R_{DSon}/R_{DSon(10\ kHz)}$ vs switching frequency

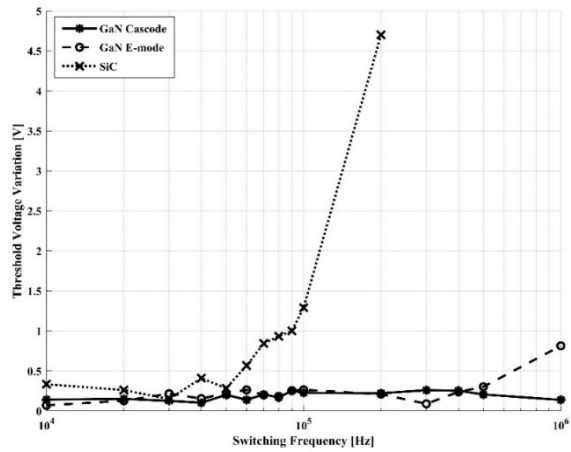

**Fig. 3.** $\Delta V_{th}$ vs switching frequency

## 4 Conclusions

The frequency behavior of the extracted figures of merit physically resides in the electron trapping and de-trapping times. For GaN devices, the electron trapping time is orders of magnitude lower than the de-trapping time [3]. Increasing the frequency, both the off-state and on-state times are reduced, but since the trapping time is much smaller, this will not affect the electron trapping but only the electron de-trapping. This leads to a bigger amount of trapped electrons at higher frequencies and so to an increase of the on-resistance and threshold voltage variation [4].

For SiC devices, during off-state, electrons tunnel out of the oxide causing a negative shift of $V_{th}$. During on-state, electrons can tunnel back into the oxide, causing a positive shift of $V_{th}$ [5] [6]. Since the device is driven with 0-15V gate voltage, the positive variation is much higher than the negative one leading to a higher positive $\Delta V_{th}$.

The present work shows the behavior of the most relevant figures of merit of GaN and SiC transistors. However, the investigation is still under progress to deepen the involved phenomena and their causes [7].

**Acknowledgments** This work has been partially financed by the EC-H2020 Project "GaN4AP" (Proposal n. 101007310 - H2020-ECSEL-2020-1-IA-two-stage).

## References

1. Mohan, N., Underland, T.M. and Robbins, W.P. Power Electronics Converter, Applications and Design, 2003, John Wiley & Sons, Inc.
2. Ortiz-Conde A. et al., Revisiting MOSFET threshold voltage extraction methods, Microelectronics Reliability, Volume 53, Issue 1, 2013, Pages 90-104, ISSN 0026-2714, https://doi.org/10.1016/j.microrel.2012.09.015.
3. J. Lei et al., Precise Extraction of Dynamic Rdson Under High Frequency and High Voltage by a Double-Diode-Isolation Method, in IEEE Journal of the Electron Devices Society, vol. 7, pp. 690-695, 2019, doi: 10.1109/JEDS.2019.2927608.
4. M. Meneghini et al., Charge Trapping in GaN Power Transistors: Challenges and Perspectives, 2021 IEEE BiCMOS and Compound Semiconductor Integrated Circuits and Technology Symposium (BCICTS), 2021, pp. 1-4, doi: 10.1109/BCICTS50416.2021.9682455.
5. K. Puschkarsky et al., Threshold voltage hysteresis in SiC MOSFETs and its impact on circuit operation, 2017 IEEE International Integrated Reliability Workshop (IIRW), 2017, pp. 1-5, doi: 10.1109/IIRW.2017.8361232.
6. A. J. Lelis et al., Basic Mechanisms of Threshold-Voltage Instability and Implications for Reliability Testing of SiC MOSFETs, in IEEE Transactions on Electron Devices, vol. 62, no. 2, pp. 316-323, Feb. 2015, doi: 10.1109/TED.2014.2356172.
7. A. Vella and G.C. Giaconia, Embedded System for GaN Devices Characterization, 2022, Master's Degree Thesis, University of Palermo